\documentclass[preprint,12pt]{elsarticle}

\usepackage{graphicx}
\usepackage{amsmath}

\usepackage{lineno}
\usepackage{subcaption}
\usepackage{amsfonts}

\usepackage{color}

\journal{Journal Name}

\begin{document}

\begin{frontmatter}


\title{Magnetohydrodynamic with Adaptively Embedded Particle-in-Cell model: MHD-AEPIC}



\author[um]{Yinsi Shou}
\author[um]{Valeriy Tenishev}
\author[um]{Yuxi Chen}
\author[um]{Gabor Toth}
\author[um,finland]{Natalia Ganushkina}

\address[um]{Department of Climate and Space Sciences and Engineering, University of Michigan}
\address[finland]{Finnish Meteorological Institute, Helsinki, Finland}

\begin{abstract}

Space plasma simulations have seen an increase in the use of magnetohydrodynamic (MHD) with embedded Particle-in-Cell (PIC) models. This combined MHD-EPIC algorithm simulates some regions of interest using the kinetic PIC method while employing the MHD description in the rest of the domain. 
The MHD models are highly efficient and their fluid descriptions are valid for most part of the computational domain, thus making large-scale global simulations feasible. 

However, in practical applications, the regions where the kinetic effects are critical can be changing, appearing, disappearing and moving in the computational domain. If a static PIC region is used, this requires a much larger PIC domain than actually needed, which can increase the computational cost dramatically. 

To address the problem, we have developed a new method that is able to dynamically change the region of the computational domain where a PIC model is applied. 
We have implemented this new MHD with Adaptively Embedded PIC (MHD-AEPIC) algorithm using the BATS-R-US Hall MHD and the Adaptive Mesh Particle Simulator (AMPS) as the semi-implicit PIC models. 
We describe the algorithm and present a test case of two merging flux ropes to demonstrate its accuracy. The implementation uses dynamic allocation/deallocation of memory and load balancing for efficient parallel execution. We evaluate the performance of MHD-AEPIC compared to MHD-EPIC and the scaling properties of the model to large number of computational cores.

\end{abstract}


\end{frontmatter}


\section{Introduction}
\label{S:intro}

Two-way coupled magnetohydrodynamic (MHD) with embedded Particle-in-Cell (PIC) models have been increasingly applied to space and planetary plasma simulations \citep{Toth:2016,Chen:2017,Chen:2019,Ma_mars:2018}. 
The MHD model can provide efficient solution based on the fluid description in a large scale global simulation, while the PIC model overcomes the limitations of the MHD model in regions where kinetic effects become critical and the fluid approximation no longer holds. 
Though the embedded PIC model is applied in a relatively small region inside the whole MHD computational domain, it can be critical to resolve the physical details on kinetic scales that go beyond the fluid approximation employed in MHD models. 
For example, Toth et al. \cite{Toth:2016} successfully applies a Hall MHD model with embedded PIC model to the magnetosphere of Ganymede. The model successfully captures the formation and evolution of a flux transfer events with similar magnetic signatures as observed by the Galileo mission.

The idea of applying coupled models to resolve different physics and achieve better efficiency has been employed in various applications. Bourgat et al. \cite{Bourgat:1996} coupled a global Navier-Stokes solver in hypersonic rarefied flows with a local kinetic solver of the Boltzmann equations in the boundary layer. Tiwari \citep{Tiwari:1998} and Degond et al. \cite{Degond:2010} proposed adaptive domain decomposition techniques for the coupling between the Boltzmann equation and the Euler equations. For plasma simulations, Sugiyama et al.\cite{Sugiyama:2007}, Daldorff et al.\cite{Daldorff:2014}, and Makwana et al.\cite{Makwana:2017} have successfully coupled MHD with PIC models, while Rieke at al.\cite{Rieke:2015} solves the Vlasov-Maxwell equations discretized on a high-dimensional grid and embed it into a 5-moment fluid model. Markidis et al.\cite{Markidis:2014} coupled the kinetic effects into the multi-fluid equations by calculating the stress tensor from the macro-particles. Holderied et al. \cite{Holderied:2021} coupled linearized, ideal MHD equations in curved, three-dimensional space, to the full-orbit Vlasov equations via a current coupling scheme, and the method conserves mass, energy, and the divergence-free magnetic field, irrespective of metric (space curvature), mesh parameters and chosen order of the scheme. To get more physically accurate results, several new methods and algorithms have been developed to ensure Gauss's law, divergence-free magnetic field and conservation of energy are satsified in PIC models \cite{Lapenta:2017,Chen:2019,campospinto2016,Kraus2017}.   Many efforts have been devoted to improving the computational efficiency of these coupled models.  One direction is to run the massively parallelized model on GPUs \cite{Rieke:2015, Shah:2016, Guo:2016, Fatemi:2017}. Another direction is to apply the kinetic region adaptively. Lautenbach and Grauer \cite{Lautenbach:2018} adaptively apply the Vlasov, 5-moment or 10-moment models to different regions according to certain criteria. In this paper, we have applied a similar idea and present a new method with an adaptive PIC region embedded into the MHD model for plasma simulations.

The new feature of the algorithm described here is its capability of changing the location of the PIC regions embedded in the computational domain of an MHD model. In application to magnetic reconnection, for example, this is very useful, since the reconnection sites are likely to move. If the PIC model is only applied to a small region, the reconnection site will move out of the region after some time and interesting phenomena are no longer captured with the kinetic model. If the static PIC region is large enough to cover all potential reconnection sites, the computational cost may become unfeasible. In this work, we propose the MHD with Adaptively Embedded PIC (MHD-AEPIC) algorithm.  The small PIC region adapts and moves with the reconnection sites, so that we can track the interesting phenomena with affordable computational resources. 

In Section \ref{S:method}, we describe the MHD-AEPIC algorithm and its implementation using the Block-Adaptive-Tree-Solar wind-Roe-Upwind-Scheme (BATS-R-US) MHD model and 
the Adaptive Mesh Particle Simulator (AMPS) 
implicit PIC model coupled through the Space Weather Modeling Framwork (SWMF). In Section \ref{S:results}, we present a test case of two merging flux ropes in a plasma moving at a constant velocity relative to the computational grid. It is shown that the new method allows the PIC region to follow the flux ropes and the results are similar (within the discretization errors) to the solution obtained with a static PIC region. The performance analysis of the new algorithm is also presented. We conclude with Section 4.

\section{Algorithm}\label{S:method}

This paper presents a new algorithm: magnetohydrodynamics with adaptively embedded particle-in-cell (MHD-AEPIC), which is an improved version of the MHD-EPIC model by \citet{Daldorff:2014}. 
Here we use the BATS-R-US code \cite{Powell:1999, Toth:2008} as the global MHD model in the whole computational domain, and the  AMPS  code \cite{Tenishev:2008} to solve the Vlasov-Maxwell equations in the embedded PIC region. The two models are two-way coupled through the Space Weather Modeling Framework (SWMF) \cite{Toth:2012swmf}.

Different from the iPIC3D code \cite{Markidis:2010} used by \citet{Daldorff:2014}, the AMPS grid is block-based. 
It divides the AMPS computational domain into multiple blocks and the same calculation or procedure is performed for each individual block. 
This feature allows AMPS to dynamically include and exclude blocks during the simulation. 
Therefore, an adaptive PIC region can be achieved by switching on and off the blocks in the PIC grid. In this section, the AMPS code and the BATS-R-US code, which provide the foundation to the MHD-AEPIC model, are described first. The new MHD-AEPIC algorithm is presented in detail in Section \ref{S:coupling}.  

\subsection{AMPS PIC model}\label{S:pic}

The Adaptive Mesh Particle Simulator (AMPS) is a fully kinetic code developed at the University of Michigan. Originally, it was designed to use the Direct Simulation Monte Carlo (DSMC) method for various comet and planetary applications. The DSMC method is one of the most frequently used approaches for solving the Boltzmann equation numerically. The simulated gas is represented by a large but finite set of macro-particles governed by the same physics laws that affect real molecules in the simulated environment. Macroscopic properties (density, velocity and temperatures) are computed by appropriately taking statistics of particle masses, locations, and velocities. 
The code has been successfully applied to rarefied atmospheres of many planetary objects, such as comet 67P/Churyumov-Gerasimenko, plumes of Enceladus and Mars’ exosphere as well \cite{Tenishev-2011-AJ,Tenishev-2013-icarus,Tenishev-2010-JGR,Lee-2015-JGR}. 
 
Recently, the PIC method has been implemented into AMPS to enable it to calculate the electric and magnetic fields self-consistently. Ions and electrons are treated as particles, while electric and magnetic fields are solved on the computational grid from the Maxwell equations. 
In particular, we have implemented the energy conserving semi-implicit method (ECSIM) first developed by Lapenta \cite{Lapenta:2017} and later improved to the Gauss-Law satisfying GL-ECSIM algorithm by Chen and T\'oth \cite{Chen:2019}. Its semi-implicit nature allows the PIC model to run on a coarser grid with a larger time step than explicit PIC methods. If periodic boundary conditions are used, the integral of the total energy, i.e. the sum of the electromagnetic field energy and particle kinetic energy, over the computational domain is conserved to the round-off error. The energy conserving property helps eliminating numerical instabilities and spurious waves that would change the total energy. The improvements by \cite{Chen:2019} ensure/improve charge conserving and further reduce the development of spurious oscillations.

ECSIM uses a staggered grid, where the electric field is defined at cell nodes, and the magnetic field is defined at cell centers. The Maxwell's equations are solved implicitly:
\begin{eqnarray}
\frac{\mathbf{B^{n+1}}-\mathbf{B^{n}}}{\Delta t} &=& -c\nabla\times\mathbf{E}^{n+\theta} \label{eqn:faraday} \\
\frac{\mathbf{E}^{n+1}-\mathbf{E}^{n}}{\Delta t}&=&c\nabla\times\mathbf{B}^{n+\theta}-4\pi\mathbf{\bar{J}}
\end{eqnarray}
where $\theta\in [0.5, 1]$  is the time centering parameter. $\mathbf{\bar{J}}$ is the predicted current density at $n+\frac{1}{2}$ time stage, and it can be expressed as a linear function of the
unknown electric electric field $\mathbf{E}^{n+\theta}$ (see equation 6, 15 in \cite{Lapenta:2017}).
The variables at the time stage $n+\theta$ can be written as linear combinations of values at the time steps $n$ and $n+1$:
\begin{eqnarray}\label{eqn:Etheta}
\mathbf{E}^{n+\theta}&=&(1-\theta)\mathbf{E}^{n}+ \theta\mathbf{E}^{n+1} \\
\mathbf{B}^{n+\theta}&=&(1-\theta)\mathbf{B}^{n}+ \theta\mathbf{B}^{n+1}
\end{eqnarray}
After rearranging the equations above and using the identity $\nabla \times \nabla \times \mathbf{E} = \nabla(\nabla \cdot \mathbf{E})-\nabla^2\mathbf{E}$, we can derive an equation with $\mathbf{E}^{n+\theta}$ as the unknown variables:
\begin{equation}\label{eqn:EField}
\mathbf{E}^{n+\theta}+\delta^2\left[\nabla\left( \nabla\cdot \mathbf{E}^{n+\theta}\right)-\nabla^2\mathbf{E}^{n+\theta} \right]=\mathbf{E}^n+\delta\left( \nabla\times \mathbf{B}^n-\frac{4\pi}{c}\mathbf{\bar{J}}\right)
\end{equation}
where $\delta=c\theta\Delta t$. After applying finite difference discretizations to the gradient and divergence operators, we obtain a linear system of equations for the discrete values of $\mathbf{E}^{n+\theta}$ at the cell nodes. The iterative generalized minimal residual method (GMRES) is used to solve the equations to obtain $\mathbf{E}^{n+\theta}$. 
Using equations~(\ref{eqn:faraday}) and (\ref{eqn:Etheta}), the magnetic field $\mathbf{B}^{n+1}$ and electric field $\mathbf{E}^{n+1}$ at the next time step can be obtained, respectively.

The position and velocity of the macro-particle are staggered in time, i.e., the particle velocity $\mathbf{v}_p$ is at the integer time stage and the location $\mathbf{x}_p$ is at the half time stage. First the velocity is pushed to time level $n+1$ by solving
\begin{equation}
\mathbf{v}_p^{n+1}=\mathbf{v}_p^{n}+\frac{q_p \Delta t}{m_p}
\left(\mathbf{E}^{n+\theta}(\mathbf{x}_p^{n+1/2})+
\frac{\mathbf{v}_p^{n}+\mathbf{v}_p^{n+1}}{2}
\times \mathbf{B}^{n}(\mathbf{x}_p^{n+1/2}) \right) 
\end{equation}
for $\mathbf{v}_p^{n+1}$ (see \cite{Lapenta:2017} for detail). 
The fields $\mathbf{E}^{n+\theta}(\mathbf{x}_p^{n+1/2})$ and $\mathbf{B}^{n}(\mathbf{x}_p^{n+1/2})$ are interpolated to the particle locations $\mathbf{x}_p^{n+1/2}$. $q_p$ and $m_p$ are charge and mass of the particle.  Finally, the particle position is updated to a preliminary new position:
\begin{equation}
 \mathbf{\tilde x}_p^{n+3/2} = \mathbf{x}_p^{n+1/2}+\Delta t \mathbf{v}_p^{n+1}  
\end{equation}
Because Gauss's law is not automatically satisfied or controlled in the original ECSIM algorithm, artificial effects can develop in long simulations. Chen and T\'oth \cite{Chen:2019} proposed several methods to reduce the error and eliminate the artificial effects. In this paper, we used the ``approximate global correction'' method  to reduce the error of $\nabla \cdot \mathbf{E}-4\pi\rho_c$ deviating from zero, where $\mathbf{E}$ is the electric field and $\rho_c$ is the net charge density in CGS units. The correction method solves the Poisson equation for the potential $\phi$ after the electric field and the particle positions are updated:
\begin{equation}\label{eqn:laplacian}
    \nabla^2\phi=\nabla \cdot \mathbf{E}^{n+1}-4\pi\tilde{\rho}_c^{n+1},
\end{equation}
where $\tilde{\rho}_c^{n+1}$ is the uncorrected net charge density interpolated in time between the net charges obtained from the particles at $\mathbf{x}_p^{n+1/2}$ and from the preliminary positions $\mathbf{\tilde x}_p^{n+3/2}$. The potential $\phi$ and net charge density $\rho_c$ are both defined at the cell center. The electrons, in general the light particles, are displaced to drive the right hand side of equation~(\ref{eqn:laplacian}) closer to zero. The displacement is
\begin{equation}
    \Delta x_p = -\frac{\epsilon}{4\pi\gamma_c\rho_{e,g}}\nabla \phi,
\end{equation}
where $\rho_{e,g}$ is the electron charge density at the closest cell center to the particle. The interpolation coefficient $\gamma_c=0.51$. $\epsilon$ can take value from 0 to 1 to avoid overshoot. It is set to 0.9 in this work. The final particle location $\mathbf{x}_p^{n+3/2}$ is
\begin{equation}
\mathbf{x}_p^{n+3/2}= \mathbf{\tilde x}_p^{n+3/2}+\Delta x_p
\end{equation}
Detailed proof and reasoning can be found in \cite{Chen:2019}. After the particles are pushed to next time step, particle moments will be calculated and stored on each cell node, which update the value of $\mathbf{\bar{J}}$ in equation~(\ref{eqn:EField}).  

The energy and charge conserving semi-implicit PIC method facilitates the coupling
between the MHD and the PIC models, since the time step in AMPS can be comparable to the time step in BATS-R-US, unlike the time step in an explicit PIC code that is limited by the stability conditions \cite{Lapenta:2017} to $\Delta t< \Delta x/c$ and the grid resolution  $\Delta x < \zeta\lambda_D$, where $c$ is the speed of light, $\lambda_D$ is the Debye length, and $\zeta$ is of order one and depends on the numerical details of the method \cite{Lapenta:2012}. In the coupled codes, the BATS-R-US MHD model provides initial and boundary conditions for AMPS, while AMPS calculates macroscopic quantities (densities, velocities, pressures and magnetic field) on the active grid blocks of the AMPS computational domain and feeds them back to BATS-R-US. The coupling process is described in detail in Section~\ref{S:coupling}. 

\subsection{BATS-R-US MHD model}
Block‐Adaptive‐Tree‐Solarwind‐Roe‐Upwind‐Scheme (BATS‐R‐US) is a flexible and highly efficient global MHD code that has been successfully applied to study plasma interactions with a wide range of planetary objects including planets, planetary moons, and comets. BATS‐R‐US also allows adaptive mesh refinement in combination with curvilinear coordinates. In the applications discussed in this paper, the code is configured to solve the two-fluid MHD equations with a separate electron pressure equation:
\begin{eqnarray}
\frac{\partial \rho}{\partial t}&+&\nabla\cdot\left( \rho \mathbf{u}\right) =0\\
\frac{\partial \rho\mathbf{u}}{\partial t}&+&\nabla\cdot \left[ \rho\mathbf{uu}+
\mathbf{I}\left(p+p_e+\frac{B^2}{2\mu_0} \right)-\frac{\mathbf{BB}}{\mu_0} \right]=0\\
\frac{\partial \mathbf{B}}{\partial t}&+&\nabla\times \mathbf{E}=0\\
\frac{\partial e}{\partial t}&+&\nabla\cdot \left[\mathbf{u}\left( \frac{1}{2}\rho u^2
+\frac{\gamma p}{\gamma-1}\right)+\mathbf{u_e}p_e+\frac{\mathbf{E}\times\mathbf{B}}{\mu_0}\right]=p_e\nabla\cdot\mathbf{u_e}
\label{eqn:energyDensity}\\
\frac{\partial p_e}{\partial t}&+&\nabla\cdot\left( p_e \mathbf{u_e}\right)=-(\gamma_e-1)p_e\nabla\cdot\mathbf{u_e}
\label{eqn:pe}
\end{eqnarray}
$\rho$, $\mathbf{u}$ and $p$ are mass density, velocity and pressure of the ion species, while $p_e$ and $\mathbf{u_e}$ are the electron pressure and electron velocity, respectively. $\gamma$ and $\gamma_e$ are the adiabatic indexes for ions and electrons, which are set to 5/3 here. $\mathbf{I}$ is the identity matrix and $\mu_0$ is vacuum permeability. The electron velocity $\mathbf{u_e}$ is
\begin{equation}
\mathbf{u_e}=\mathbf{u}-\frac{\mathbf{J}}{|q_e| n_e}
\end{equation}
where $n_e$ and $q_e$ are the electron number density and the electron charge, respectively,
and the current density $\mathbf{J}$ is obtained from
\begin{equation}
\mathbf{J}=\frac{1}{\mu_0}\nabla\times\mathbf{B}.
\end{equation}
The electric field $\mathbf{E}$ is computed from the generalized Ohm's law:
\begin{equation}
\mathbf{E}= -\mathbf{u_e}\times\mathbf{B}-\frac{\nabla p_e}{|q_e| n_e}
\end{equation}
The total energy density $e$ includes the total ion energy density and magnetic energy density:
\begin{equation}
e=\frac{p}{\gamma-1}+\frac{\rho u^2}{2}+\frac{B^2}{2\mu_0}
\end{equation}
Note that the electron thermal energy $p_e/(\gamma_e-1)$ is not included in $e$, which results in the source term on the right-hand side of equation~(\ref{eqn:energyDensity}). This does not violate the conservation of total energy $e+p_e/(\gamma_e-1)$ as it cancels out with the right hand side source term of the electron pressure equation~(\ref{eqn:pe}).

Hyperbolic/parabolic cleaning method by Dedner et al. \cite{Dedner:2001} is used in combination with the eight‐wave scheme \cite{Powell:1999} to control the numerical divergence of the magnetic field. For pure MHD and Hall‐MHD simulations, the hyperbolic/parabolic cleaning is not necessary, but for MHD‐EPIC the divergence error cannot propagate through the PIC region and can accumulate at the boundary of the PIC region, since AMPS does not use the eight‐wave scheme. The hyperbolic/parabolic cleaning solves the issue, because it can dissipate the divergence error in all directions as found by \cite{Toth:2017}.

\subsection{Adaptively embedded PIC region}\label{S:coupling}

\begin{figure}[ht]
\centering\includegraphics[width=1.0\linewidth]{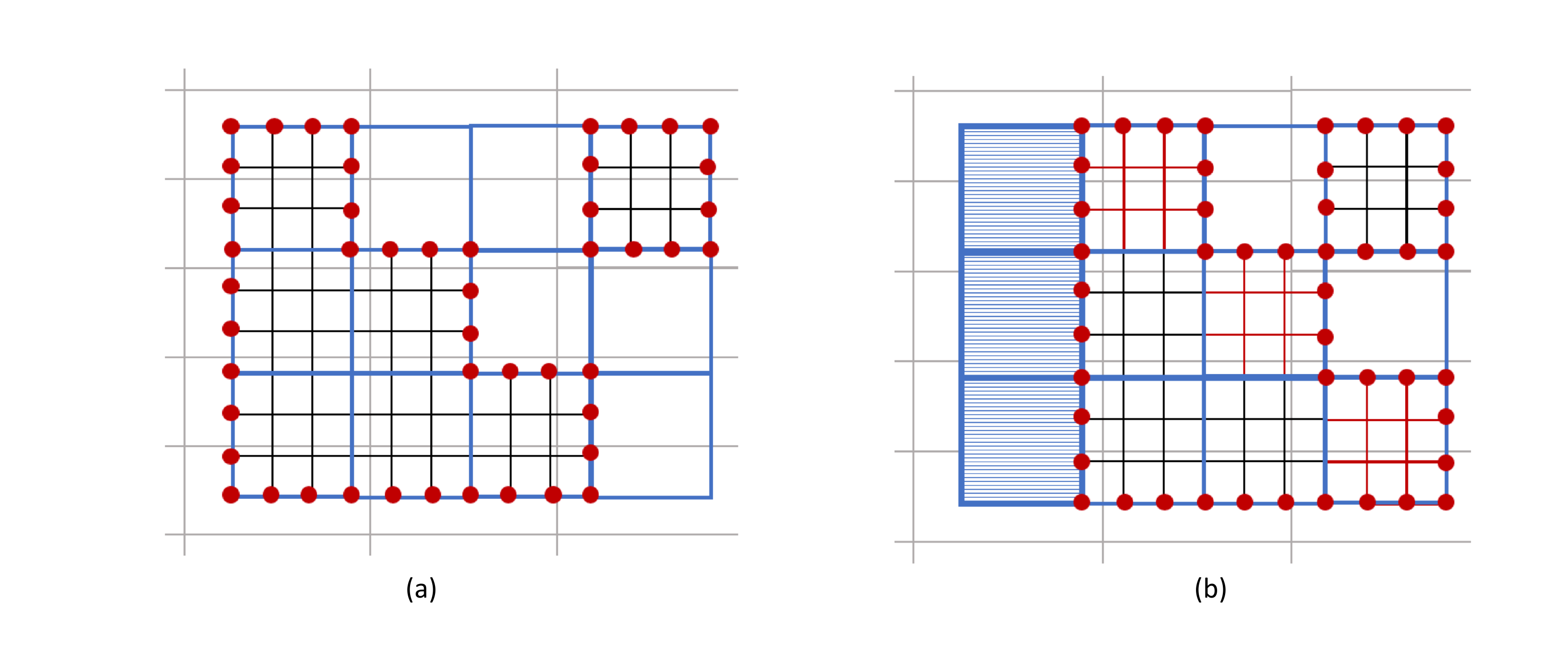}
\caption{Schematics of the MHD-AEPIC algorithm.
Panels (a) and (b) show a PIC domain consisting of $4\times3$ grid blocks before and after adaptation with the underlying MHD grid represented by the gray lines. The inactive PIC blocks are empty, while the active blocks contain a mesh of $3\times3$ grid cells. The boundary cell nodes of the active PIC regions, indicated by solid red circles, are interpolated from the MHD solution. Panel (a) contains two discontinuous active PIC regions. In this example, the region on the left moves to the right due to adaptation. Panel (b) shows the newly allocated grid blocks with red cells initialized from the MHD solution, while the deallocated blocks are covered by blue horizontal stripes.}
\label{fig:schematic}
\end{figure}

AMPS uses a block-based grid to discretize the computational domain.
Fig.~\ref{fig:schematic} shows an example how the dynamic allocation and deallocation of PIC blocks allow adapting the embedded PIC region. The AMPS code first creates a box-shaped PIC grid divided into grid blocks. In the example, the PIC grid has 4 by 3 blocks, which are denoted by squares with blue edges. White squares with blue edges represent inactive and unallocated blocks, while active blocks are shown with a black grid representing the cell edges.  The active blocks allocate memory for cell nodes and cell centers, which store variables such as electric and magnetic fields needed for PIC calculations. There can be several disconnected active PIC regions in the PIC grid at the same time. Panel (b) shows two PIC regions, one consisting of 6 blocks and the other one a single block. The red solid circles in the figure show the boundary cell nodes of the adaptive PIC regions. The cells containing any of the boundary cell nodes are boundary cells. The electric field $\mathbf{E}$ at the boundary cell nodes are fixed to values interpolated from the MHD grid, which is denoted by gray color. Similarly, at the boundary cell centers, the mass density $\rho_s$, the bulk velocities $\mathbf{u}_s$, the pressure $p_s$ for species $s$ and magnetic field $\mathbf{B}$ are also set to interpolated MHD values. The particles in boundary cells will first be removed and generated based on the acquired MHD values. The thermal velocity of particles of the species $s$ follows the Maxwellian distribution, i.e. the probability distribution function of the thermal velocity $a_{s,i}$ in the $i=x,y,z$ direction is $f(a_{s,i})\sim \exp(-\frac{a^2_{s,i}}{kT_s/m_s})$, where $k$ is the Boltzmann constant and $T_s$ is the temperature for species $s$. The thermal velocity for one macro-particle in three directions can be generated by random variables $R_1$, $R_2$, $R_3$ and $R_4$ with uniform distributions in $(0,1]$, 
\begin{eqnarray}
a_{s,x}=\sqrt{\frac{p_s}{\rho_s}}\sqrt{-2\ln{R_1}}\cos(2\pi R_2),\\
a_{s,y}=\sqrt{\frac{p_s}{\rho_s}}\sqrt{-2\ln{R_1}}\sin(2\pi R_2),\\
a_{s,z}=\sqrt{\frac{p_s}{\rho_s}}\sqrt{-2\ln{R_3}}\cos(2\pi R_4).
\end{eqnarray}
The velocity $\mathbf{v}_s$ of a macro-particle is the sum of the bulk velocity $\mathbf{u}_s$ and the thermal velocity: $\mathbf{v}_s=\mathbf{u}_s+\mathbf{a}_s $. Next, the macro-particle is created at a random location $\mathbf{x}_s$ in the cell with a uniform distribution. Finally, the weight of the macro-particle $w_s$ is obtained by $w_s= \rho_s(\mathbf{x}_s) V/(m_sN_s)$, where $\rho_s(\mathbf{x}_s)$ is the density linearly interpolated to the particle location, $V$ is the cell volume and $N_s$ is the number of macro-particles in the cell. The weight $w_s$ equals to the number of protons or electrons represented by one macro-particle. Since the number of macro-particles for a certain species in a grid cell is given by $N_s$, the macro-particle weights can be computed from the total mass in the cell volume $\rho_s V$ divided by $m_s N_s$, where $m_s$ is the mass of a real particle, for example a proton.

If the adaptive PIC region does not change for a period of time, the physical quantities in the boundary cells in AMPS code are only updated when coupling or communication between AMPS and BATS-R-US happens. During the two-way coupling process, AMPS also provides feedback to BATS-R-US. The MHD values on the MHD grid covered by the active PIC region are overwritten by values from the AMPS code. AMPS first computes integrations of particle mass density, momentum and velocity tensor in each cell and stores them on cell nodes. The density at a cell node at $\mathbf{x}_n$ is obtained as
\begin{equation}
\rho_s(\mathbf{x}_n)= \sum_p b_1(\mathbf{x}_n-\mathbf{x}_p) \frac{w_p m_s}{V}, 
\end{equation}
where the sum is over the particles in surrounding cells. $\mathbf{x}_p$ and $w_p$ are the location and weight of the particle. $b_1(\mathbf{x}_n-\mathbf{x}_p)$ is the first order spline function to account for the contribution of the particle to the node. The MHD mass density is obtained by summing over species: $\rho = \sum_s \rho_s$.

Similarly, the momentum $\rho_s\mathbf{u}_s$ and the kinetic energy tensor $\mathbf{K}_s$ are computed as follows:
\begin{eqnarray}
\rho_s\mathbf{u}_s(\mathbf{x}_n) &=&
\sum_p b_1(\mathbf{x}_n-\mathbf{x}_p) \frac{w_p m_s \mathbf{v}_s}{V},\\ 
\mathbf{K}_s(\mathbf{x}_n) &=&
\sum_p b_1(\mathbf{x}_n-\mathbf{x}_p) 
\frac{w_p m_s \mathbf{v}_p\mathbf{v}_p}{V}.
\end{eqnarray}
The MHD momentum is $\rho\mathbf{u} = \sum_s \rho_s \mathbf{u}_s$.
The pressure tensor elements can be computed by subtracting the kinetic energy density tensor of the bulk flow:
\begin{equation}
P_{s,ij}= K_{s,ij} - 
\frac{(\rho_s u_{s,i})(\rho_s u_{s,j})}{\rho_s}.
\end{equation}
Since the pressure is  isotropic in the MHD model used here, the scalar ion and electron pressures are obtained from the trace of the corresponding kinetic energy tensors as  $p_s(\mathbf{x}_n) = \sum_i P_{s,ii}$/3. Then the same quantities of the MHD model (i.e. mass density, momentum, ion pressure, electron pressure and magnetic field) at the MHD cell centers that are covered by the active PIC regions are replaced with the values linearly interpolated from the cell nodes in AMPS. 

Panels (a) and (b) in Fig.~\ref{fig:schematic} show the change of the adaptive PIC region. Three active blocks in (a) are deallocated in (b), which are denoted by blue squares with horizontal stripes. The particles in these blocks are deleted and the memory that is used to store particle and field data is released. The blocks with red cells inside in panel (b) are newly allocated active blocks. The electric and magnetic fields in the new blocks are initialized from interpolated MHD values on the gray MHD grid, while particles in the cells of the new blocks are created with particle weights and velocities based on the  MHD model as described earlier in this section.  After each adaptation, load re-balancing is performed in AMPS to ensure that every MPI process has equal or similar number of active PIC blocks.

In summary the MHD-AEPIC algorithm consists of the following steps:

\begin{enumerate}

\item The MHD model is solved globally on the grey grid and the PIC model is solved in the active blocks denoted by the black grid in Fig.1.  

\item When the coupling time is reached, the PIC region is adapted based on the current solution. 

\item The mass densities, velocities, pressures and field values are updated in boundary cells of the active PIC blocks and inside the newly activated PIC blocks.

\item Old particles in the boundary cells are removed and new particles are generated based on the latest MHD values interpolated to the PIC cells.

\item The MHD variables are replaced with the magnetic field and the interpolated moments of the PIC solution in the grid cells that are covered by the active PIC blocks at step 1. Note this step is executed in the MHD model and can run in parallel with step 2 to 4.

\item A new cycle is started from step 1.

\end{enumerate}

\section{Numerical Tests}\label{S:results}
\subsection{Merging flux ropes test}

\begin{figure}[ht]
\centering\includegraphics[width=1.0\linewidth]{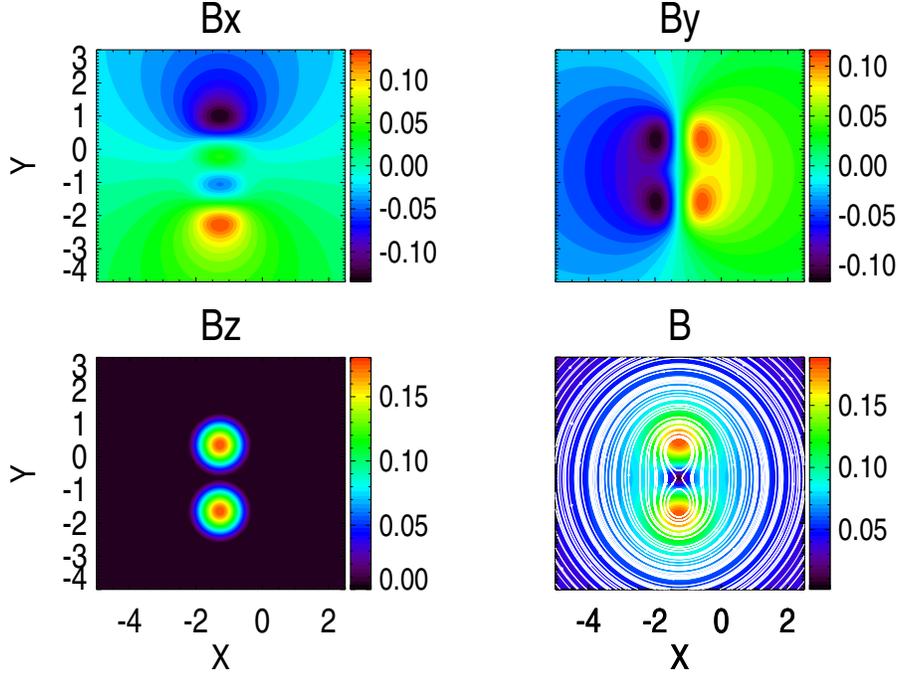}
\caption{The initial condition of the magnetic field for the two flux ropes. The magnetic field lines are shown in white in the fourth panel.
}\label{fig:b_init}

\end{figure}

We use the test case of two merging flux ropes to demonstrate how the new method works. 
The units are normalized so that the speed of light $c=1$, the magnetic permeability $\mu_0=1$, the elementary charge
$|q_e|=1$ and the proton mass $m_p=1$.
The computational domain of the MHD code is $x\in [-16,16]$, $y\in [-16,16]$ and $z\in [-0.125,0.125]$
with floating boundary conditions. 
The initial mass density is $\rho=100$, the ion pressure is $p=2.5\times 10^{-3}$ and the electron pressure is $p_e=5\times 10^{-4}$ uniformly over the whole domain.
The initial velocity is set to 
$\mathbf{u} = (0.01, 0.005, 0)$, so that the magnetic structures move across the domain and the adaptivity of the embedded PIC regions can be properly tested. 
This bulk speed is less than the maximum Alfv\'en speed 0.02 and the speed of light $c=1$. 

Initially, two force-free identical flux ropes are placed next to each other parallel with the $z$ axis.
The current density profile used in \citet{stainer2013} can be written in the cylindrical coordinates $(r,\phi,z)$ centered around the axis of the flux rope: $J_r=0$ and 
\begin{equation}
J_z(r) =
\begin{cases} 
j_m\left[1-(r/w)^2\right]^2, & \text{if $r \le w$}  \\
0, & \text{if $r > w$}
\end{cases}
\end{equation}
where $j_m$ is the maximum current density in the center of the flux rope and $w$ is the radius of the flux rope. $J_\phi$ is much more complicated and is not displayed here. As the flux rope is force-free, the current density is parallel to the magnetic field.
As magnetic fields are the fundamental variables in BATS-R-US and needed to set up the initial condition of the simulation, they are obtained from the current density and the force-free condition as $B_r=0$ and
\begin{eqnarray}
B_{\phi}(r) &\!=\!&
\begin{cases} 
j_m\left(\frac{r}{2} + \frac{r^5}{6w^4} - \frac{r^3}{2w^2}\right), & \text{if r $\le$ w}  \\
\frac{w}{r}B_{\phi}(w), & \text{if $r>w$}
\end{cases} \\
B_z(r) &\!=\!& 
\begin{cases}
j_m\sqrt{\frac{47}{360}w^2 - \frac{1}{2}r^2 + \frac{3}{4}\frac{r^4}{w^2} - \frac{5}{9}\frac{r^6}{w^4}   + \frac{5}{24}\frac{r^8}{w^6} - \frac{1}{30}\frac{r^{10}}{w^8}}, & \text{if $r \le w$} \\
0, & \text{if $r>w$}
\end{cases}
\end{eqnarray}
$j_m=0.5$ is used in this work so that the plasma-beta is the order of one near the flux ropes. The axes of the two flux ropes are at $x=-1.28$, $y=-1.64$ and at $x=-1.28$, $y=0.36$, respectively. Both flux ropes have radii $w=1$. The initial magnetic fields are shown in Fig.~\ref{fig:b_init}. Because of the parallel currents along their central axes, the two flux ropes will attract and move towards each other. In the merging process, magnetic reconnection takes place, when two flux ropes are compressed together with anti-parallel magnetic fields. The process simulated in this work is a fully kinetic collisionless reconnection. Introductions on the subject of magnetic reconnection can be found in plasma textbooks, such as \cite{gurnett_book_2005}.

\begin{figure}[ht]
\centering\includegraphics[width=1.1\linewidth, trim={6cm 0 6cm 0}, clip]{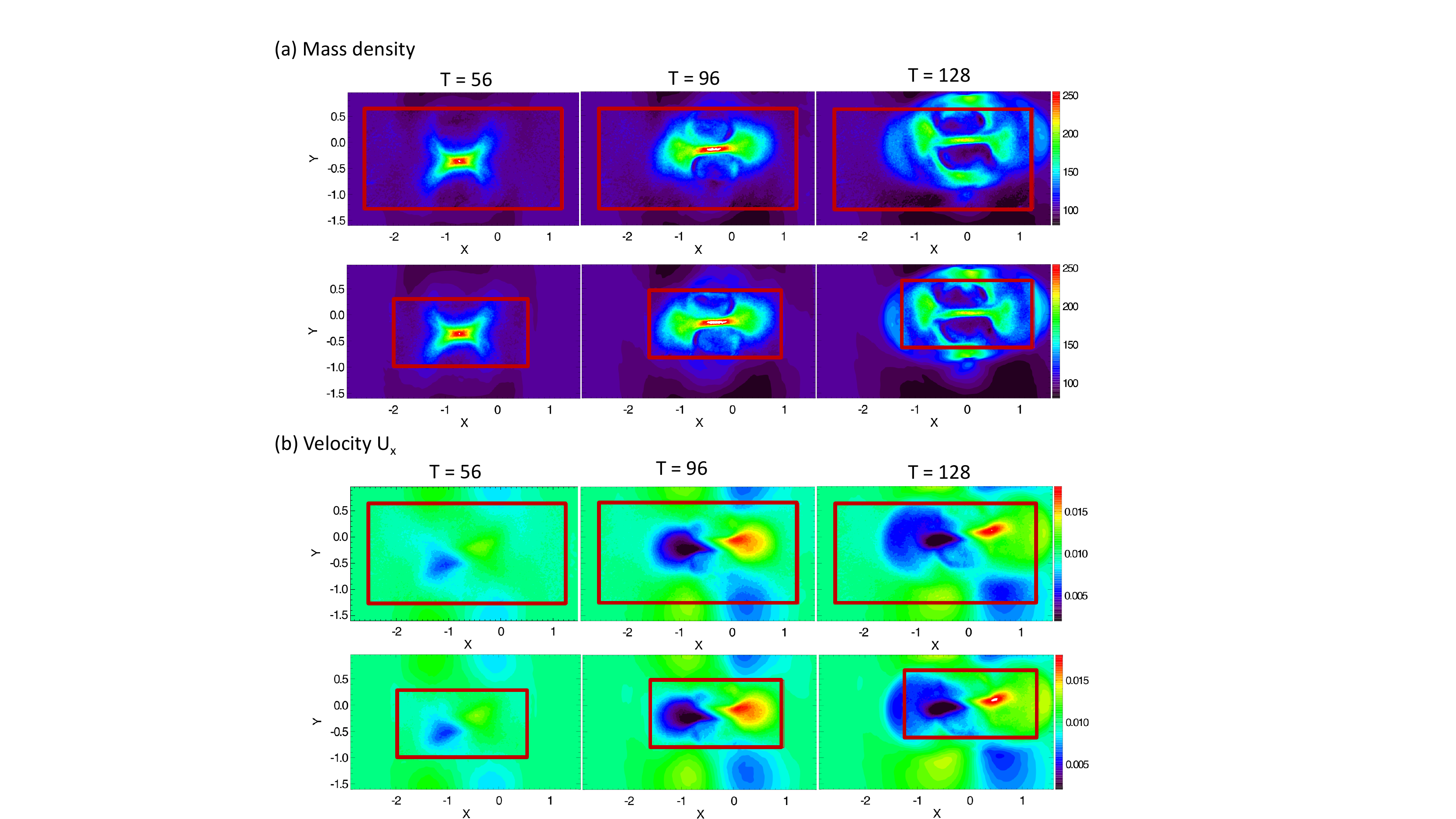}
\caption{The plasma density, the x component of the plasma velocity from the MHD model output at $t=56$, 96 and 128 from the left to the right column. The  upper two rows show the plasma density and the bottom ones show the the x component of the plasma velocity. In each pair of rows, the upper one shows the static reference results and the lower one shows the adaptive simulation result. The PIC region in each plot is denoted by a red box.}\label{fig:gm_output}
\end{figure}

The MHD grid is refined near the center of the domain at $x\in [-3,1.5]$, $y\in [-2,1]$ and $z\in [-0.1,0.1]$. The cell size in this central region is $\Delta x_{MHD} = \Delta y_{MHD}= \Delta z_{MHD}=1/64$. The cell resolution of the one layer transition region around the refined central region is $1/32$ in each direction. The rest of the domain has a cell resolution of $1/16$.  A Courant-Friedrichs-Lewy (CFL) number of 0.6 is used in the MHD model to set the time step.
The MHD-EPIC model has a static PIC region with $x\in [-2.56,1.28]$, $y\in [-1.28,0.64]$, $z\in [-0.08,0.08]$. The adaptive MHD-AEPIC simulation has a smaller moving PIC region initially at $x\in [-2.56,0]$, $y\in [-1.28,0]$, $z\in [-0.08,0.08]$. This region is moved  with the initial background velocity of 
$\mathbf{u}=(0.01, 0.005, 0)$
so that the merging part of the two flux ropes remain near the center of the active PIC region. 

The cell size in the PIC model is $\Delta x_{PIC}=\Delta y_{PIC}=\Delta z_{PIC}=0.01$. Given the normalization of proton mass per charge and the speed of light to unity, the ion inertial length is $\rho^{-0.5} = 0.1$ that is well resolved. We set the ion-electron mass ratio $m_i/m_e=100$, so the electron skin depth is $(\rho\, m_i/m_e)^{-0.5} = 0.01$ that is marginally resolved. To ensure that the CFL number $u\Delta t_{PIC}/\Delta x_{PIC}$ is about 0.2 with $u$ as the maximum characteristic electron velocity, we set the PIC time step as $\Delta t_{PIC}=0.005$. The species used in the PIC model are protons and electrons. $N_s=100$ particles are created for each species when generating particles in a grid cell. Each block in the PIC model has 8$\times$8$\times$4 cells for the flux rope test. There are 2048 blocks in the MHD-AEPIC simulation and 4608 blocks in the MHD-EPIC simulation. The coupling time between the MHD and PIC model is enforced to be the same as $\Delta t_{PIC}$. The resulting time steps of the MHD and PIC model for the runs in this paper are the same, but this is not required in general.

\begin{figure}[ht]
\centering\includegraphics[width=1.0\linewidth, trim={0cm 10cm 0cm 0}, clip]{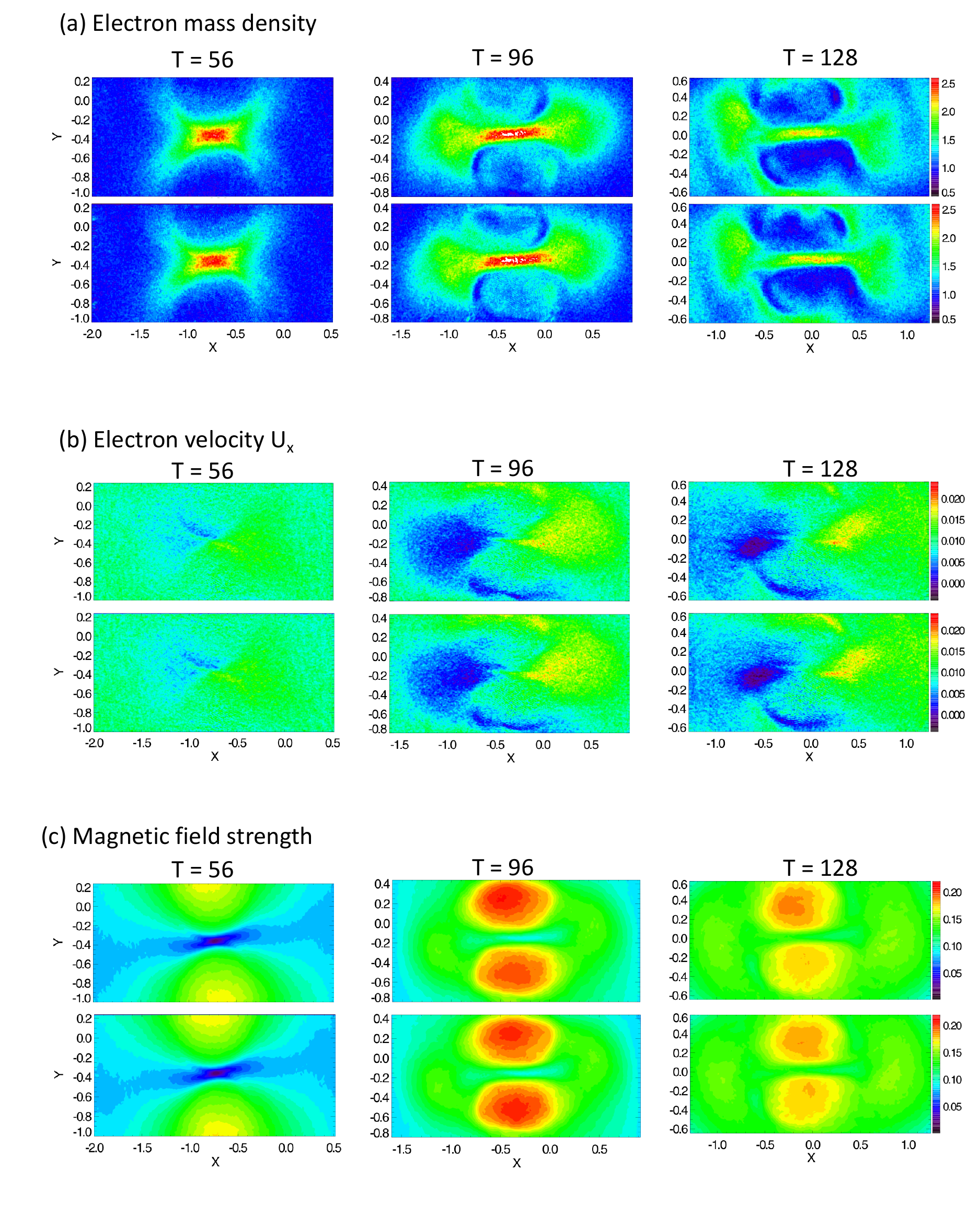}
\caption{The electron density, the x component of the electron velocity, and the magnetic field magnitude from the PIC model output at $t=56$, 96 and 128. In each pair of rows, the upper row shows the reference static simulation result and the lower row shows the adaptive simulation result.}\label{fig:pc_output}
\end{figure}

Figure~\ref{fig:gm_output} shows the results from the MHD model. Plasma density shown in the upper two rows and the x component of the plasma velocity in the lower two rows at simulation times $t=56$, 96 and 128. For each time the upper row gives the static MHD-EPIC simulation results and the lower row shows the MHD-AEPIC simulation results. The PIC regions are denoted by red boxes. It can be seen from the figure that the static PIC simulation has a larger PIC region, while in the  adaptive simulation the red box is moving in the direction from lower left to upper right.
Fig.~\ref{fig:pc_output} compares the results from the PIC model for the static MHD-EPIC and adaptive MHD-AEPIC simulations at the same times. The figure displays the electron density, $x$ component of the electron velocity and magnetic field magnitude at the location of the adaptively embedded PIC region at the given time. 
Both the MHD and PIC model output show the process of the two flux ropes merging. 
The plasma from the top part flows in negative y-direction, colliding with the upward flowing plasma from the bottom part carrying oppositely oriented magnetic fields. Around simulation time $t=56$, the X-line starts forming and electrons get accelerated in the x-direction. 
After the onset, the reconnection process continues. At $t=96$ and $t=128$, the two flux ropes are getting closer and the magnetic field magnitude diminishes as a result of magnetic reconnection transforming magnetic energy into kinetic energy.

Comparing the static and adaptively embedded simulations, the results look almost identical to each other, except for the regions that are solved by PIC in the static simulation but are computed by the MHD model in the adaptive simulation. For example, the arc below the reconnection site in the center of the mass density plots at time $T=128$ in Fig.~\ref{fig:gm_output} is more enhanced in the adaptive case than in the static case. Nevertheless, the similar results prove that the new adaptive embedded PIC region method has the capability of preserving all signatures and physical processes
happening in a static PIC region. 

The computational times are compared for runs on Frontera \cite{Frontera} on 1024 Intel Xeon Platinum 8280 CPU cores at a nominal clock rate of 2.7GHz. The codes are compiled by Intel compilers with an optimization level  of three. The BAT-S-RUS code uses about 98 seconds of wall clock time in the MHD-AEPIC simulation and 112 seconds in MHD-EPIC simulation. The AMPS code uses about 775 seconds in the MHD-AEPIC simulation and 1549 seconds in the MHD-EPIC simulation, which indicates the new method is more
efficient. Table~\ref{tab:breakdownFluxRope} shows the comparison of the breakdown of CPU times used by the two models. The major algorithmic steps are the field solver, particle mover, Gauss' law correction, and collecting moments as described in Section~\ref{S:pic}. The adaptation step is unique to the MHD-AEPIC model and is described in Section~\ref{S:coupling}. The times shown in the table are the maximum times of all MPI processes. If the implementation of the algorithm was perfect and the test case was ideal, the ratio of the simulation times of the adaptive and static PIC models is expected to be the same as the ratio of the computational volumes, which is 4/9 in this test. However, the ratio of the simulation times is about 1/2, indicating that the efficiency is not ideal. One reason that can be seen from Fig.~\ref{fig:pc_output} is that the particles are mostly concentrated in the adaptive region, and the ratio of particle numbers most of the time is higher than $4/9\approx 0.44$. For instance, the ratio of particle numbers is about 0.52 at $t=128$. 

\begin{table}
\centering
\caption{Breakdown of CPU times (in seconds) for the flux rope test.}\label{tab:breakdownFluxRope}
\begin{tabular}{lrr}
\hline
    & EPIC & AEPIC  \\
\hline
Field solver       & 460  &  229  \\
Particle mover     & 192   & 77  \\
Gauss' law correction    & 564  &  235 \\
Collecting moments & 284  & 116  \\
Adaptation            & N/A  & 57 \\
\hline
Total               & 1500   & 714 \\
\hline
\end{tabular}

\end{table}

\subsection{Weak scaling test}

\begin{figure}[ht]
\centering\includegraphics[width=1.0\linewidth]{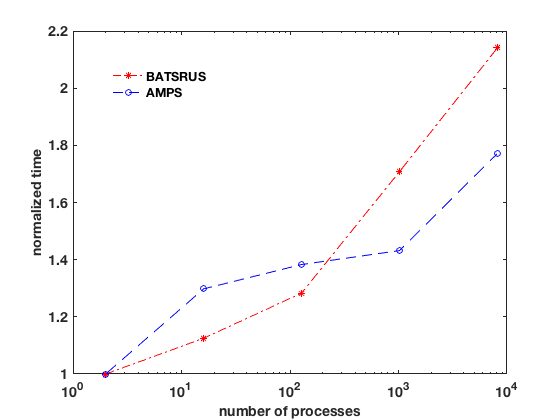}
\caption{The weak scaling curve of the algorithm. The x-axis shows the number of CPU cores or MPI processes, the y-axis shows the average time spent for one PIC iteration normalized by the time of the 2-process run. The red asterisks and blue circles represent the timings for the BATS-R-US and AMPS codes, respectively.}\label{fig:scaling}
\end{figure}

A weak scaling test is performed to evaluate the parallel performance of the implementation of the new MHD-AEPIC algorithm. The simulation used for the weak scaling test is a slightly modified version of the flux rope merging test presented in the previous subsection.  The following adjustments are made to accommodate the need of the weak scaling test.  A smaller computational domain $x\in [-4,4]$, $y\in [-4,4]$, $z\in [-4,4]$ with a uniform mesh is applied to the MHD model. The box for the PIC model has lengths of 2.56, 1.28 and 0.64 in the $x$, $y$ and $z$ directions, respectively. A PIC block has $16\times$8$\times4$ cells and an MHD block has $8\times$8$\times8$ cells for the scaling test. A series of tests are run on 2, 16, 128, 1024 and 8192 CPU cores or MPI processes. The grid resolutions are chosen for each test so that there are 4 PIC blocks and 32 MHD blocks on each processor.

For the run with 2 processes, the cell resolution in each direction is $0.25$ for the MHD model and is $0.08$ for the PIC model. The cell resolution is halved for the run on 8 times as many processes, so that the cell numbers and computational load are roughly the same for each process for one iteration. The time step for the PIC model is fixed to 0.1 for the 2-process run and is reduced proportionally to the cell resolution to keep the CFL condition the same. For the MHD models, the CFL number is fixed to 0.6 for all the runs.
The simulations are run to $t=36$. 

Fig.~\ref{fig:scaling} shows the average wall clock time per time step normalized by the 2-processor runs as a function of the number of MPI processes. The red asterisks represent the BATS-R-US timings and the blue circles represent the AMPS timings. The average time for the 2-process run spent 0.12$s$ on BATS-R-US and 0.47$s$ on AMPS. It can be seen as the number of processes increases, more time is used for both the BATSRUS code and AMPS code. The average wall clock time on 8192 cores are about 2.2 and 1.8 times longer than the 2-core run for the BATSRUS and AMPS codes, respectively. We also want to point out that the time spent on the adaptation part is insignificant. The 8192-process run uses a smaller cell size and smaller time steps in the PIC region, so it involves more frequent dynamic allocation and deallocation of blocks in the same simulation time than other runs. The percentage of CPU time spent on adaptation increases as more processes are used. The 8192-process run uses about 5\% of the AMPS computational time on adaptation.

\subsection{Grid convergence study}
A grid convergence study is presented in this section to show how the error decreases with grid refinement. We use a similar setup as in the weak scaling test and perform $N=5$ simulations with grid resolutions $\Delta x_{PIC,i}=0.04, 0.02, 0.01, 0.005$, and 0.0025 for $i=1\ldots N$, respectively. The ratios $\Delta x_{MHD,i}/\Delta x_{PIC,i}=3.125$ and $\Delta t_i/\Delta x_{PIC,i}=1.25$ are kept constant. As mentioned before, the electron skin depth is about 0.01, so the finest grid with $\Delta x_{PIC,N}=0.0025$ is expected to resolve the reconnection physics best and is used as the reference solution.  To reduce the computational cost, a 2D version of the MHD-AEPIC model is utilized, where only one grid cell is used in the $z$ direction both in the MHD and PIC models and a periodic boundary condition is imposed on the particles in the z-direction. In the convergence study, it is guaranteed that all simulations use the same adaptive PIC region at the same time. Other parameters are the same as in the weak scaling test. 

\begin{figure}[ht]
\centering\includegraphics[width=1.0\linewidth]{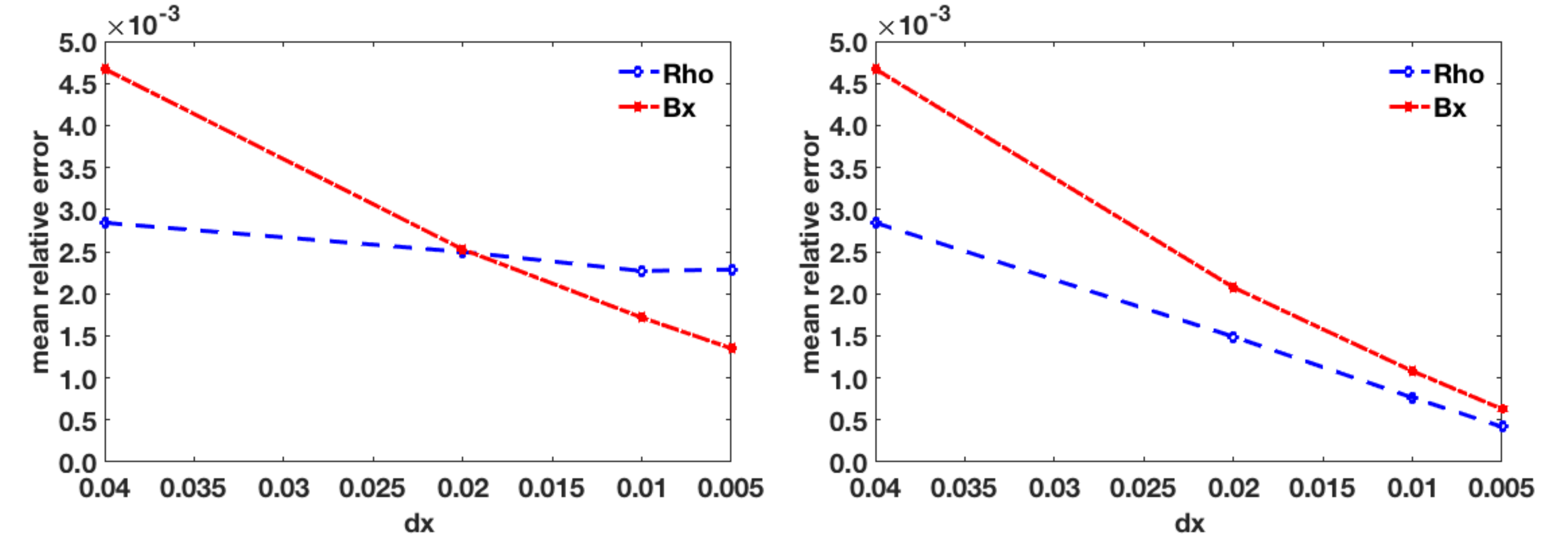}
\caption{
Grid convergence test results obtained with two different approaches. The y-axis is the mean relative error compared to the reference solution and the x-axis is the grid cell size in the PIC model. The left panel shows the errors computed for the grid cells of the given simulation. The right panel, on the other hand, shows the errors calculated after the solutions of the given simulation as well as the reference solution
are integrated over the grid cells of the coarsest grid with $\Delta x=0.04$. The red asterisks and blue circles represent the mean relative errors for the mass density $\rho$ and the x component of magnetic field $B_x$, respectively.
}\label{fig:convergence}
\end{figure}

The results of the 5 runs at time $t=36$ are analyzed using two different approaches for the convergence study. In the first approach, the reference solution $Q_N$ is averaged over the grid cells of size $\Delta x_{PIC,i}$ for run $i$ and compared to the solution $Q_i$.
A mean relative error $E_i$ for each run $i=1\ldots(N\!-\!1)$ is obtained by averaging $|Q_i\,-\!<\!\!Q_N\!\!>_i\!\!|/|\!\!<\!\!Q_N\!\!>_i\!\!|$ over all the grid cells of run $i$. The left panel of Fig.~\ref{fig:convergence} shows the mean relative errors obtained with this approach. The blue line shows the mean relative error for the mass density $\rho$ and the red line represents the error for the magnetic field component $B_x$. The mean relative error for $B_x$ has a steeper slope than the mass density error with increasing grid resolution. In fact, the density error stops decreasing at $\Delta x_{PIC}$=0.01. This is due to the statistical noise caused by the finite number of macro-particles $N_p$, which is roughly constant per grid cell for each simulation. The error due to the statistical noise is proportional to $1/\sqrt{N_p}$, which is not decreasing with the grid refinement.

In the second approach, we average both the reference solution $Q_N$ and the simulation result $Q_i$ over the coarsest grid cells with $\Delta x_{PIC,1}$=0.04 so we calculate the error of the quantity integrated over a fixed volume. The mean relative error for each simulation $i=1\ldots(N\!-\!1)$ is obtained by averaging $|\!\!<\!\!Q_i\!\!>_1-<\!\!Q_N\!\!>_1\!\!|/|\!\!<\!\!Q_N\!\!>_1\!\!|$ over all the grid cells of the coarsest grid. The right panel of Fig.~\ref{fig:convergence} shows the mean relative errors obtained on the coarsest grid with this approach. The errors for both quantities drop approximately linearly as the resolution gets higher. The statistical noise for mass density in the coarsest cell is still proportional to $1/\sqrt{N_{p,i}}$, but here $N_{p,i}$ represents the number of particles of simulation $i$ in the volume of the coarsest cell, and thus $N_{p,i}$ is inversely proportional to $\Delta x_{PIC,i}^2$. As a result, the statistical noise in the second approach is $O(\Delta x_{PIC})$. The other discretization errors are at least first order accurate, so the overall convergence rate is first order. In 3D, one may obtain a convergence rate of 1.5 using integrals over 3D volumes.

In general, calculating errors for integrated quantities over fixed size volumes seems to be the proper approach for doing grid convergence studies for PIC simulations. 

\section{Conclusion}
In this paper, we have presented a new MHD-AEPIC algorithm that dynamically adapts the PIC region embedded into an MHD model. It can be used in the cases where the phenomena of interest  that need to be resolved by kinetic models are not stationary and appear, disappear or simply move in the computational domain. If such dynamic phenomenon occurs in a large area, the MHD-EPIC model employing a static PIC region will be inefficient. 
The new adaptive MHD-AEPIC method can track the phenomenon with smaller PIC regions and requires less computational resources. 

A numerical test of two merging flux ropes is used to demonstrate the capability of the new method to track the moving magnetic reconnection site and preserve its major features. MHD-AEPIC is also shown to be more efficient than the MHD-EPIC model. A weak scaling test is performed to demonstrate the efficiency of the implementation on a large number of CPU cores.
In the future, this method can be widely used in space plasma simulations. For example, Earth magnetosphere simulations have moving magnetic reconnection sites both at the dayside magnetopause and in the magnetotail. The moving PIC region can follow the reconnection sites without requiring very large PIC regions in the simulation and can provide a more efficient model of the evolving magnetosphere. 

\bigskip
\noindent\textbf{Acknowledgments}

Support for this work was provided by grant 80NSSC17K0681 from the NASA Living with a Star Program, grant 80NSSC20K0854 from the NASA Solar System Workings Program and grant PHY-1513379 from NSF INSPIRE program. The authors also acknowledge the Texas Advanced Computing Center (TACC) at The University of Texas at Austin for providing high performance computing resources on the Frontera supercomputer that have contributed to the research results reported within this paper.

\bibliographystyle{elsarticle-num-names}
\bibliography{csem.bib}

\end{document}